# *Inbetween*: Visual Selection in Parametric Design

The Typographic Case Study

RONY GINOSAR and AMIT ZORAN

The Hebrew University and The Bezalel Academy of Arts and Design, Jerusalem, Israel

rony.ginosar@mail.huji.ac.il , zoran@cs.huji.ac.il

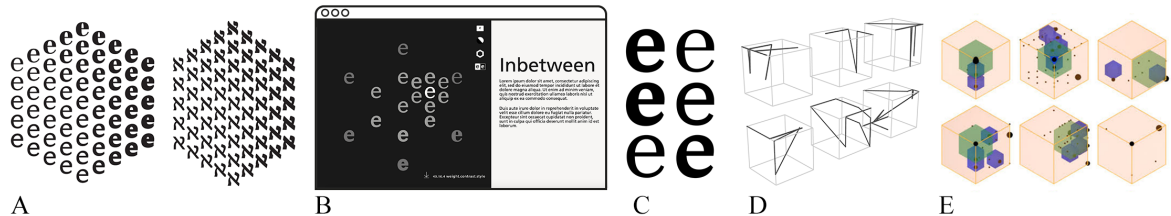

Figure 1: *Inbetween* typographic selection. (A) Typographic parametric space (Latin and Hebrew). (B) Screenshot of Inbetween visual catalog. (C) Typeface instances selected by a designer for different tasks. (D) Traces of the selection process within the parametric space. (E) Interest areas in the parametric space are highlighted by light green and dark purple volumes, start and end typeface instances are marked by larger black dots.

The act of selection plays a leading role in the design process and in the definition of personal style. This work introduces visual selection catalogs into parametric design environments. A two-fold contribution is presented: (i) guidelines for construction of a minimal-bias visual selection catalog from a parametric space, and (ii) *Inbetween*, a catalog for a parametric typeface that adheres to the guidelines, allows for font selection from a continuous design space, and enables the investigation of personal style. A user study conducted among graphic designers, revealed self-coherent characteristics in selection patterns, and a high correlation in selection patterns *within* tasks. These findings suggest that such patterns reflect personal user styles, formalizing the style selection process as traversals of decision trees. Together, our guidelines and catalog aid in making visual selection a key building block in the digital creation process and validate selection processes as a measure of personal style.

**CCS Concepts** • Human-centered computing ~ Human computer interaction (HCI) ~ Interactive systems and tools • Human-centered computing ~ Interaction design ~ Interaction design process and methods

## 1 INTRODUCTION

In a creative process, the personal style of the creator is manifested through a process of visual selection, [30]. Visual catalogs are widely used in Human-Computer Interaction (HCI), from online shopping to creative work environments (e.g., Adobe Photoshop brushes). However, digital selection tools often do not encourage visual selection *per se*, creating a detachment from the visual creative process. This effect may worsen as parametric tools are introduced into design, complicating the selection process due to an intangible infinite range of options. As in cases of daily selections, too many options make it harder to choose [60]. Visual selection is an artistic action, and, as with style, it is complex and dynamic, not solely reflected by the final choice, hence the need for process-centered tools [17].

Studying the selection process may aid in creating better design tools which can reduce the amount of data required to make a decision. Current digital tools typically overlook the visual aspects of typeface selection. For example, traditional dropdown menus of typeface names lack sufficient visual information for typeface selection. By means of toolmaking, we studied the nature of the visual selection process. We aimed to learn users' personal styles through their unique visual selection patterns and to then apply the conclusions towards creation of parametric catalogs. As a case study, we concentrated on the field of typography, constituting of letters that form a well-defined set of visuals. We constructed a typographic selection catalog for a custom-made parametric typeface. Twenty-one practicing graphic designers with varying levels of experience, performed font selection tasks out of a visual catalog. Our data revealed self-coherent characteristics in selection patterns (i.e., a user presenting similar selection patterns while preforming different tasks), regardless of the user's level of expertise. This suggests that such patterns reflect the user's personal style, opening up a path towards the definition of selection processes as a measure of personal style.

This work contributes (i) catalog-creation guidelines, derived from an in-depth review [25] of factors that influence visual selection, aspiring to minimal visual bias (*Minimal-Bias*) and (ii) *Inbetween*, a visual selection catalog adhering to the guidelines, enabling the investigation of user style (https://ronyginosar.github.io/parametricSpecimen).

We open this work by reviewing visual selection factors and selection aspects in design and HCI. We continue by describing our *Inbetween* catalog implementation and presenting a user study and its findings. We close with a discussion proposed future directions and conclusions.

## 2 BACKGROUND & RELATED WORK

### 2.1 Style and Selection

Personal style is manifested in the set of choices a maker makes [30], which can be characterized by static and dynamic aspects [18]. The former refers to finished works of an artist, while the latter refers to the process of creation, including the artist's set of decisions applied throughout the creation process. In a digital context, style research usually addresses two main topics: (i) algorithmic analysis of style (e.g., stylization [24] or art history research [21]) and (ii) personal style, using digital tools (e.g., [83]). Within the digital realm, and specifically in HCI, style is defined as the visual representation of the artist's work and not the process that produced it [18].

Selection is a constant component of our day, including the decision of what to eat and wear. Many selections are made from visual options, involving pattern recognition [47,79]. Visual search within options is led by the salience of objects and task-related factors [37]. Thus, the visual *design* of the set of choices can influence the speed and ease of visual processing during selection [40], and even the choice strategies applied in the selection process [66].

Too many choices exhaust us and lead us to avoid making selections altogether [60]. The vastness of options and information available to users cause overload [23,67], as often experienced while shopping online ("tyranny of choice" [67]). Since overload is not a constant state during everyday situations [65], understanding when it occurs is crucial for creating minimal-overload visual selection catalogs.

We wish to make parametric tools accessible to creators implementing their style. A visual process is introduced into digital design tools, to render them more intuitive and selection-oriented. To this end, we curate the set of options to select from and the manner of their display, while creating as little overload as possible.

## 2.2 Selection in HCI

Visual catalogs are increasingly present in our daily technology interactions, from online browsing to creative work environments (e.g., a color selection panel). Visual catalogs proposed [39] to familiarize craftsmen with new technologies, were implemented through a standard digital concept of 'presets.' A need for easy manipulation of modular elements was indicated [39]. A pre-generated task-specific catalog [20] enabled the fine-tuning of a selected object, and a later pre-generated catalog [26] opened a discussion of what the requirements from a catalog are. Following these discussions, we focus on visual selection in the creative process and how it relates to style and expertise.

Visual selection is an essential component of the creative task. It relies on the expertise of an expert, is frequently an initial step of the creative exploration process, and often involves multiple iterations [80]. Traditionally, the creator visually selects building blocks for their creative process (e.g., a carpenter selects a wood log for a desk). Various manual crafts use 'visual grammar,' indicating possible connections and options, such as digital snapping parts [64]. Visual cues also facilitate the learning of new techniques and environments, a notion that can be translated into digital tool-making. However, most digital tools prioritize smooth user experience over retaining visual selection. Consequently, many digital tools may appear '*cold*,' which can be alienating for some craftsmen.

Visual images and categorization are extensively used in online browsing. Visuals capture initial attention and facilitate comparison, with text providing a second layer of information [59]. Moreover, visuals are processed as a whole, faster than textual and verbal, and visuals enable items to be filtered out during the process [29,73]. An object-centered interface using categorization, filtering, organizational structure, and visual information descriptions, assists the user in processing a large amount of data [42].

In our work, we follow these facilitating lines, enabling visual selection with inherent categorization and filtering tools. In contrast to existing visual catalogs, we exploit the observation that visual selection is routine for the creator and devote our attention to the selection process in designing our tool.

## 2.3 Selection in Parametric Design

The use of sliders in user interfaces is widespread. Sliders are approachable and accessible, allowing rapid onboarding to new interfaces. While there are cases when a slider is appropriate (e.g., volume adaptation), it is often used as a default for range selection, calling for 'blind selection' (i.e., moving back and forth until the right variation is found) [28]. In parametric design user interfaces, multiple sliders are commonly used to allocate numerous parameters, accumulating an infinite amount of optional combinations, resulting in choice overload.

In this work, we wish to improve selection involving multiple parameters by enabling an intuitive rather than a blind search. *Inbetween* enables shape-based selection and direct manipulation, defining parametric artifacts as digital building blocks, with which the user can select and interact.

## 2.4 Selection in Typography

Text and typography are ever-present in our daily lives and are fundamental building blocks in visual communications [33]. By selecting specific typefaces, using a combination of experience and skill [32], the designer communicates distinct purposes and rhetorical information to a text [10,11].

Historically, typefaces[1] were selected out of a printed catalog, which visually presented the typeface qualities. In the digital scope, dropdown menus are the main typographic selection tool. Typeface selection occurs out of a long list of typeface *names,* which fail to present their visual properties, prompting selection based on pre-acquaintance [7]. With the shift to web and web-fonts, type foundries now present their catalogs in dedicated websites, using sliders and dropdown menus as interactive evaluation tools (e.g., myfonts.com). However, these selection platforms draw the creator away from visual selection, due to various screen space and technology constraints.

Research regarding typography primarily concerns legibility and readability [2,3,6,12]. HCI-focused research studies text as an information delivery system affecting user experience [32,41,51,58], decision-making process [55] (e.g., user agreements in social media [1]) and ease of technology usage (e.g., driving response times [63]). HCI typographic research proposes automated font creation or selection for casual users by semantic definition [56], attribute selection [77] and interpolation of exiting fonts [14].

In contrast, we investigate expert (rather than casual) user selection from a pre-defined parametric design space. The expert user selects an instance out of the space that has been carefully crafted by the typeface designer. Such instance selection is made to match specific expert user needs. Based on this typography case-study, we consider the construction of tool creation in parametric design in general, and reintroduce visual selection, proposing a novel solution to address the digital constraints. Providing designers with a tool to visually explore a design space, fine-tune typefaces, and even tailor familiar ones to their changing needs, may broaden their selection and inspire them within the design process.

## 2.5 Parametric Typography

The notion of a typographic space first evolved during the 20th century [8]. The Noordzij Cube [53,54] (Figure 2A) is an early description of a typographic parameter space, spanning over three dimensions (i.e., parameters); principal classic and modern typefaces inhabit its corners. The cube is an interesting case study in the parametrization of a discrete space, successful in drawing axes between separate, well-known instances.

Variable typefaces (a.k.a. OpenType Font Variations) developed from the need to efficiently deliver web fonts [35]. As a side benefit, this novel technology supports previously unavailable parametric typefaces, spanning a design space of fonts, and facilitates typeface customization. Variable typeface catalogs (e.g., v-fonts.com and play.typedetail.com) consist of sliders, each presenting one parameter of the typeface. As above, slider-based interactions suffer of 'blind selection' and overload. However, most such websites exhibit gamified interactions of the font (e.g., recursive.design) or visual mappings of existing fonts (e.g., fontmap.ideo.com), rather than present a working tool. We introduce a novel visual selection tool to be used in the workflow of designers. Its novelty lies in the interaction it allows between the user and the parametric design space. Existing display methods of parametric and typographic elements were modified in accordance with the conclusions reached from an in-depth review of factors influencing visual selection [25]. All this facilitates visual selection as a means to study *style*.

---

[1] *Typeface* is the design of the letter, e.g. Helvetica. *Font* is the instance file, e.g. Helvetica Bold.

## 3 *INBETWEEN*: IMPLEMENTING A VISUAL CATALOG FOR PARAMETRIC DESIGN

Current selection tools that enable visual selection of design elements, generally allow selection out of a static finite set. We offer a new approach, presenting a catalog of an entire design space and gradually reveal alternatives based on user interest as input. An entire space, rather than samples of the space, is provided, giving the user better control over the display. The display dynamically adapts itself according to user focus, allowing continuous, real-time shape adjustment, rather than selection from a discrete and finite set.

The presented visual catalog encapsulates a parametric design space engine in its foundation [26], incorporating a tinkering approach into the parametric design environment [64]. The catalog allows zooming-in to areas of interest and exploration of the design space that is spanned in front of the designer [82], triggering interactive and modular visual selection [39], followed by fine-tuning of parametric design space instances [20].

Re-introduction of visual selection catalogs to the creative process can facilitate the digital process for creatives, specifically visual creators. Increasing the intuitiveness of the design space concept drives exploration of the space and, consequently, exploration of the approach and its possibilities.

Direct interaction with a visual shape is an intuitive action, inducing exploration, and inspiring the creative user. Parametric interaction can be roughly divided into two approaches: (i) formal handling of an equation that receives parameter inputs, and (ii) direct shape-'kneading,' visual manipulation of the parametric-sourced artifact. Following review of user-study observations and historical surveys of the development of creative selection tools, we believe that the latter is more intuitive to creatives. By increasing intuitiveness, we can strengthen the creator-creation bond, especially in the digital-tool scope. Moreover, we promote exploration of the design space, which is enabled by parametric technologies, largely uncharted territories to many designers and creators.

### 3.1 Defining Our Parametric Space

The parametric space built for our catalog was a Hebrew implementation of the Noordzij Cube (Figure 2D), created in collaboration with a typeface designer. Using the new variable typeface technology (Section 2.5), each letter was expressed as an independent three-dimensional parametric space (e.g., Figure 2C). Thus, a selection of an instance of the typeface was a selection from within the parametric space. On a typographic note, thus far, there is no equivalent typeface in Latin[2].

---

[2] The figures in this paper were manually translated to English for the sake of presentation. See Appendix for Hebrew interface screenshots.

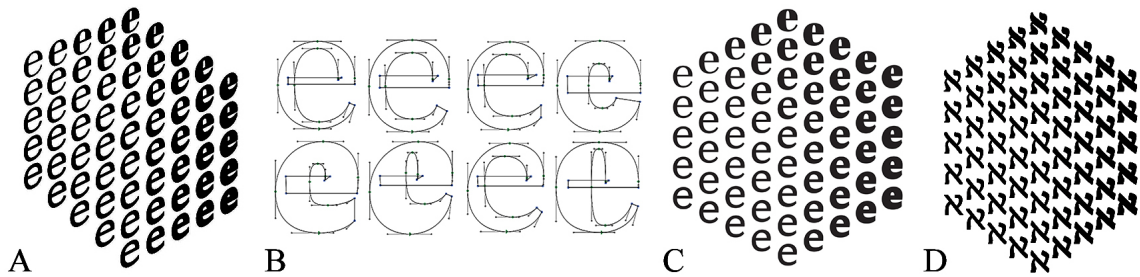

Figure 2: Creating the design space. (A) The Noordzij Cube, The Stroke, 2009, Hyphen Press (With permission). (B) Control points on extreme instances for the parametric space of 'e', produced using Glyphs (glyphsapp.com). (C,D) The resulting typographic parametric space for the letters 'e' and 'aleph' (Hebrew 'a').

### 3.2 Guidelines for Construction of a Minimal-Bias Catalog

We propose a *minimal-bias* visual catalog targeting an expert audience [36]. The catalog presents the selection process as a set of sequential subproblems, allowing the user to expand or narrow their size, essentially enabling self-management of exposure and problem resolution, and displaying a growing number of options with each subproblem. The proposed catalog has no default starting option; rather, it offers an initial partition into labeled categories. This dynamic partition, shifting between alternative and attribute comparison, alongside the use of categories to further refine the set, requires user input and engagement.

Our selection catalog is constructed to allow for user input-driven [13] shifts in attention and method [5,49] . The user initially reviews alternatives, and then, by zooming-in on an area of interest, the user discriminates and eliminates options, arriving at a phase of relative comparisons of alternatives [17]. Side-by-side visual presentation of the alternatives [34] allows for less reliance on an internal reference that would otherwise be deployed [74,75]. The display of familiar items, alongside novel alternatives [57,61,62] further aids overload reduction.

The screening process commences by showing high-level categories of the design space. Doing so reduces overload and avoids *default bias* [19,45]: The categories act as an *a priori* preference to presenting initial instances [16,31,48]. Next, zoom-based navigation within the display prompts the user to indicate desired levels of different attributes, acting both as filters and as fine-tuning tools toward selection of an instance [36]. The sequential [15,72] display of expanding [45] subproblems lowers complexity appearance [19,57,68], ensuring that the presented variations vary on the attribute most relevant to the user [57]. Zoom-based navigation lets the user control the increase in both the instance appearance and set resolution, both in level and timing [19,68,71], as the instances grow in similarity [22,43,44,46].

Partition into subproblems is a natural procedure in selection, and involves screening options to form a consideration set [27,52,78]. The use of subproblems enables presentation of complex, niche, and extreme options, also allowing for their selection [38]. Subset size is consistent with rational models and cultural habits [19]. Conventional-size notions are deployed in the assortment selection field: a small set consists of 5-7 items, while a large set includes 21-30 items. Each zoom-in enhancement adds a subproblem to the display, consisting of an even number of options to prevent a passive 'middle option' selection. Depriving an 'easy selection' leads to attribute comparison in small sets of complex alternatives [15]. The design space partition is based on immediately observable, objective, and collectively exhaustive partitions [9], implementing domain-specific jargon categories [50].

Finally, display format and space influence assortment evaluation and search behavior [4,36,70]. They should be carefully constructed, keeping a clean interface, while tending to task-unrelated factors, like font size [55].

### 3.3 Interface

We present *Inbetween*, a prototype visual catalog, implementing the aforementioned *minimal-bias* guidelines. *Inbetween* was designed to facilitate our investigation of user selection (Sections 4, 5). The workflow in *Inbetween* (Figure 3) starts with selection of a category for initial exploration, one of two opposite instances. By hovering over instances of the typeface, adjacent instances appear, sharing features with the hovered-over instance (Figure 3C). Simultaneously, hovered-over instances change the display font of the editable example text (Figure 3B). Zooming-in on an instance adds its neighbors to the display for further exploration, with three zoom levels available. Clicking on an instance opens a slider fine-tuning window before downloading the achieved instance (Figure 3D). (Note: Figure 3 is a translated version of the interface, see Appendix for Hebrew interface screenshots).

The initial visual selection offers two opposite variations, acting as an exploration starting point (Figure 3A). Each option represents a pair of common opposite high-level typographic categories, allowing for a simple selection between four high-level categories. Either choice brings the user to the main display (Figure 3B), with the selected instance appearing at the center.

Six neighboring variations appear around the initial instance, reachable, via different manipulations, from the starting instance and from one another. A hexagonal grid is used to arrange the instance display (Figure 3E), providing for a clutter-free arrangement of the 3D space on a 2D display. Inspired by the Noordzij Cube, a hex-grid enabled the simultaneous display of all variations that can be achieved from a specific instance, keeping the spatial order in each subspace and helping orientation.

Hovering over any instance reveals its neighboring instances, with the distance corresponding to the current zoom level. Each newly revealed neighbor can be seen as a mid-way variation between the focused instance and its currently displayed neighbor. In other words, each zoom-in adds more instances to the display, and these instances grow in similarity (Figure 5B). The zoom action adds the newly revealed neighbor and shifts the display center toward the zoomed area, with further adjustments available with a drag action. Clicking on any instance opens and updates the fine-tuning window to present the selected instance, showing sliders corresponding to the three axes of the typeface (Figure 3D). Each instance is identified by unique coordinates within the space, shown at the bottom of the screen, below the fine-tuning window, together with a three-word descriptor corresponding to the three axes, and with a download arrow for the current instance (Figure 3B.5). Fine-tuning updates the example-text font and coordinates. The interface is tailored for expert users, equipped with tools needed for a full evaluation of the considered typeface, clean of extraneous visuals and features (Figure 3B.3-4).

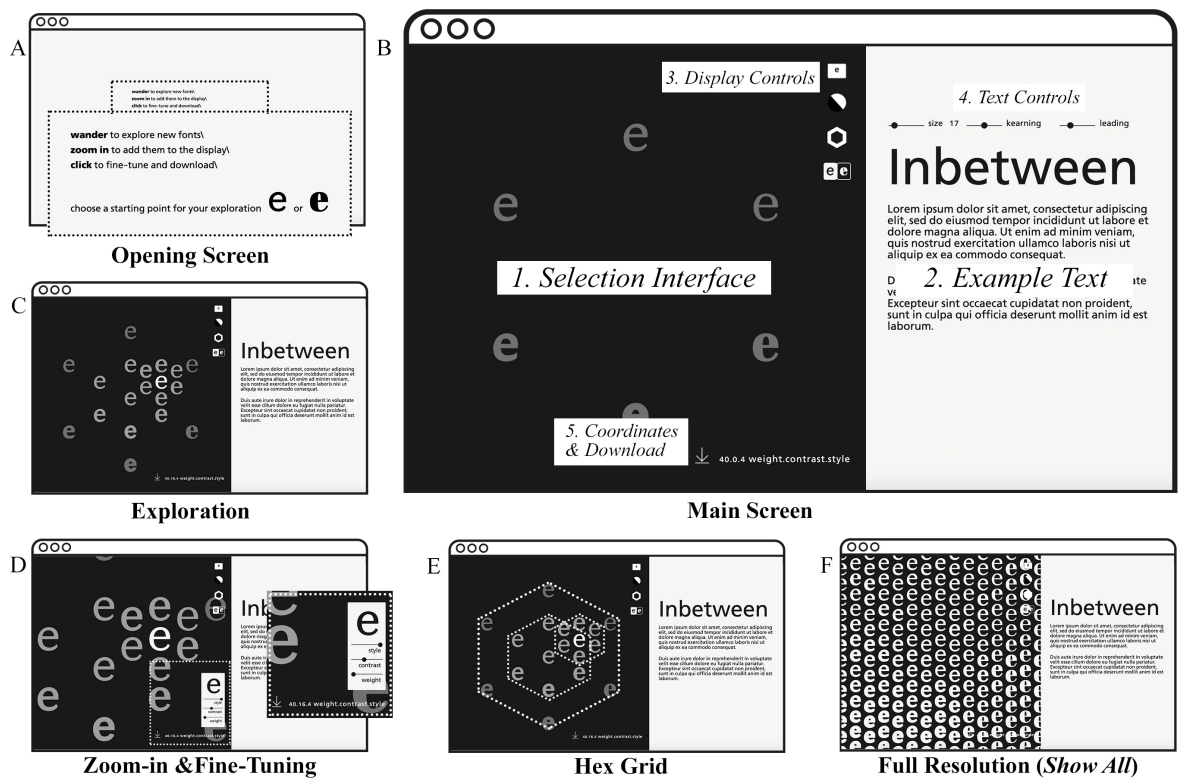

Figure 3: *Inbetween* Workflow & Interface. (A) Opening screen, selecting a starting point for exploration from high-level categories. (B1) The main screen of the *Inbetween* interface. (B2) Editable Example Text. (B3) Display control icons (top-to-bottom): the text box allows change of the letter or word displayed in the main selection area; the Contrast icon switches between the B/W display colors; the Show All Instances button (hexagon) enabled further control; the Starting point toggle switches the middle instance to the opposite option and as a result the entire instance display is also switched. Typographic evaluation: (B4) Hovering over the Example Text shows sliders for control of Text Size, Letter Spacing (Kerning), and Line Spacing (Leading). (B5) Instance Coordinates and Download button (arrow). (C) The design space is explored by hovering and zoom-in actions, which comprise the selection process. (D) Zooming-in adds variation neighbors to the hovered-over instance, at radii corresponding to zoom level. Clicking on an instance opens a fine-tuning window to refine the selection. (E) A Hex-grid-based display allows for simultaneous display, while keeping the spatial order for orientation. (F) The Show All Instances display. Note: Figure 3 is a translated version of the interface, see Appendix for Hebrew interface screenshots.

## 4 USER STUDY

The purpose of this study was to examine visual selection patterns out of catalogs, as a means to characterize *style* (Section 2). To learn about the visual selection process, we relied on the visual investigation facilitated by *Inbetween*. *Inbetween* enables identification of individual user styles by visually analyzing their selection patterns, an analysis that is not possible to conduct with slider-based interfaces. The user study targeted practicing graphic designers, asking them to complete six different font selection tasks for predefined purposes. The selections were observed along two dimensions: per user, while performing all six tasks, and per task across all users. We hypothesized that patterns and trends could be identified in the decision-making and selection process, along each of the two dimensions, especially with increasing years of experience. A secondary goal was to obtain usability feedback for the *Inbetween* interface.

## 4.1 Participants

Twenty-one participants took part in the user study (11 female, 10 male), as well as another twelve in several rounds of pre-studies. All participants were graphic designers who select fonts on a daily basis. Participants' expertise ranged from first-year visual communication students (i.e., 'Novices') to twenty years in the design profession (i.e., 'Experts'). Participant ages ranged from 23 to 37 years (avg. 30 years). All participants were native Hebrew speakers. Participants were not compensated for their time. Sessions took place in-person during work hours, in their respective offices.

## 4.2 Procedure

Participants were given six selection tasks (Figure 4C). In each task, the user was given a design assignment and selected a font from the design space to use (i.e., download) in that assignment. Defining a purpose for each selection enabled grounding the process to a common daily task [69,76]. Each font selection was performed twice; once using the proposed *Inbetween* interface (Figure 4A), and once using a control interface (Figure 4B, [84]). Divided randomly, half of the participants started their experience using the control interface, performing all six tasks, and then repeated them on to the *Inbetween* interface. The other half started with the *Inbetween* interface and then moved on to the control interface. In total, each participant completed 12 selection tasks. Each participant performed the first six tasks in random order; the same order was applied for the second set of six tasks. After each selection, the user was asked to reset the interface (i.e., refresh the web page), starting each selection process using the same initial visual environment.

The control interface is an example of the standard tools used today for parametric font interaction. The font is initially arbitrarily presented with all parameters at minimal values. Alongside the font are sliders to control the font parameters. Both the control interface and *Inbetween* presented the same font with the same (three) parameters.

The users were told that the purpose of the study was to evaluate alternative interfaces for using parametric fonts. In order to neutralize selection bias, The users were not told that their selection process was the focus of our study. The sessions commenced with an explanation of the session context, along with an introduction to the Variable Typefaces technology. Before each set of selection tasks, the user was given an explanation and demonstration of the respective interface. The user was then given practice time to use the interface until they felt they were ready to start the tasks.

In the context of collection of *Inbetween* usability feedback, participants were asked to articulate their thoughts about the selection process, their satisfaction with their choices, their experience, and how they felt the two interfaces compared with each other [84].

Both interfaces featured our typeface developed for this study and contained generally similar functionalities. All sessions were held on a MacBook Pro 15'' 2015. Participants could navigate the screen either with a mouse or a trackpad. For each user selection, the selection process for *Inbetween* and the control interface were recorded. Figure 5A shows the detailed recorded data for *Inbetween*: the Start instance of the exploration, the end instance (the downloaded font), the exploration process path - the process of traversing the decision tree (also in Figure 1D), points of interest, and the zoom-in areas of interest (also in Figure 5B).

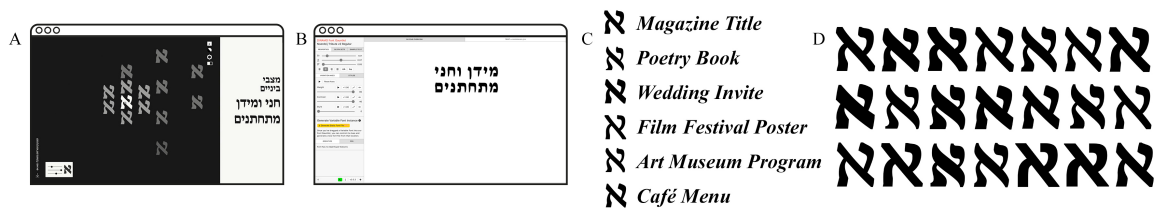

Figure 4: User Study. *Inbetween* (A) and the control interface (B). Both screenshots were taken at the end of the same selection task (Wedding Invite) of a specific user, presenting the user's final selection within each interface. Tasks and font selections from the user study: (C) User 1 selections, accompanied by the task descriptions (top to bottom). (D) Café Menu task (task #6) of all 21 users (user 1 at top left, LTR order).

## 4.3  Findings

We wish to learn about user *selections*: their personal preferences and styles, patterns of decision making, and approaches to multi-dimensional search. We hoped to conduct a quantitative analysis of these rather qualitative features. Taking a visual-centered approach, we traced the traversal of the decision tree, performed by each user while performing a task (Figure 1D, 5A). The traces were then compared and contrasted with each other. Subsequently, the tree-traversals were visually clustered. In other words, we examine the visual selection process as traversals of decision trees. This visual approach enabled us to offer insights and draw conclusions that were undiscoverable with more conventional statistical analysis.

Figure 5C displays the features of selection processes executed using our *Inbetween* catalog prototype, the semantics of which are described in Figures 5A and 5B. The 21 columns represent participants, sorted by years of experience, from left to right (U1 was a first-year graphic design student; U21 had 20 years of experience). The six rows represent the six font selection tasks: Each row is a single-themed task, disregarding the order in which the participant performed it. Each selection task is represented by a light orange hexagon, representing the design space. The hexagons contain light green and dark purple smaller hexagons, corresponding in size to different zoom levels (Figure 5B), located according to different interest areas. At the far right, we show an overlaid representation of each *task* for all *users*. Similarly, an overlaid representation of each *user* for all *tasks* is shown at the bottom.

To investigate user style and its manifestation in the selection process, we examined the selection patterns (traversal traces) of each user. Recall that, as the selection process is a unique signature of style, we expected to discover a user's *self-coherent* pattern.

***Selection Patterns.*** One-third of the participants presented a recurring pattern, consistent across different tasks (Figure 5E). This included 'non-selection' and 'minimal interaction' patterns, with the former indicative of users who primarily selected from the initially given options rather than zooming-in to interest areas (Figure 5E, orange hexagons not containing inner hexagons), and the latter referring to users who zoomed-in once and then only used sliders (Figure 5C, U8, U11, U17). One user (U12) used different search patterns and examined different areas for each task. Three users repeated patterns in different areas for each task (U2, U9, U16), as opposed to U12.

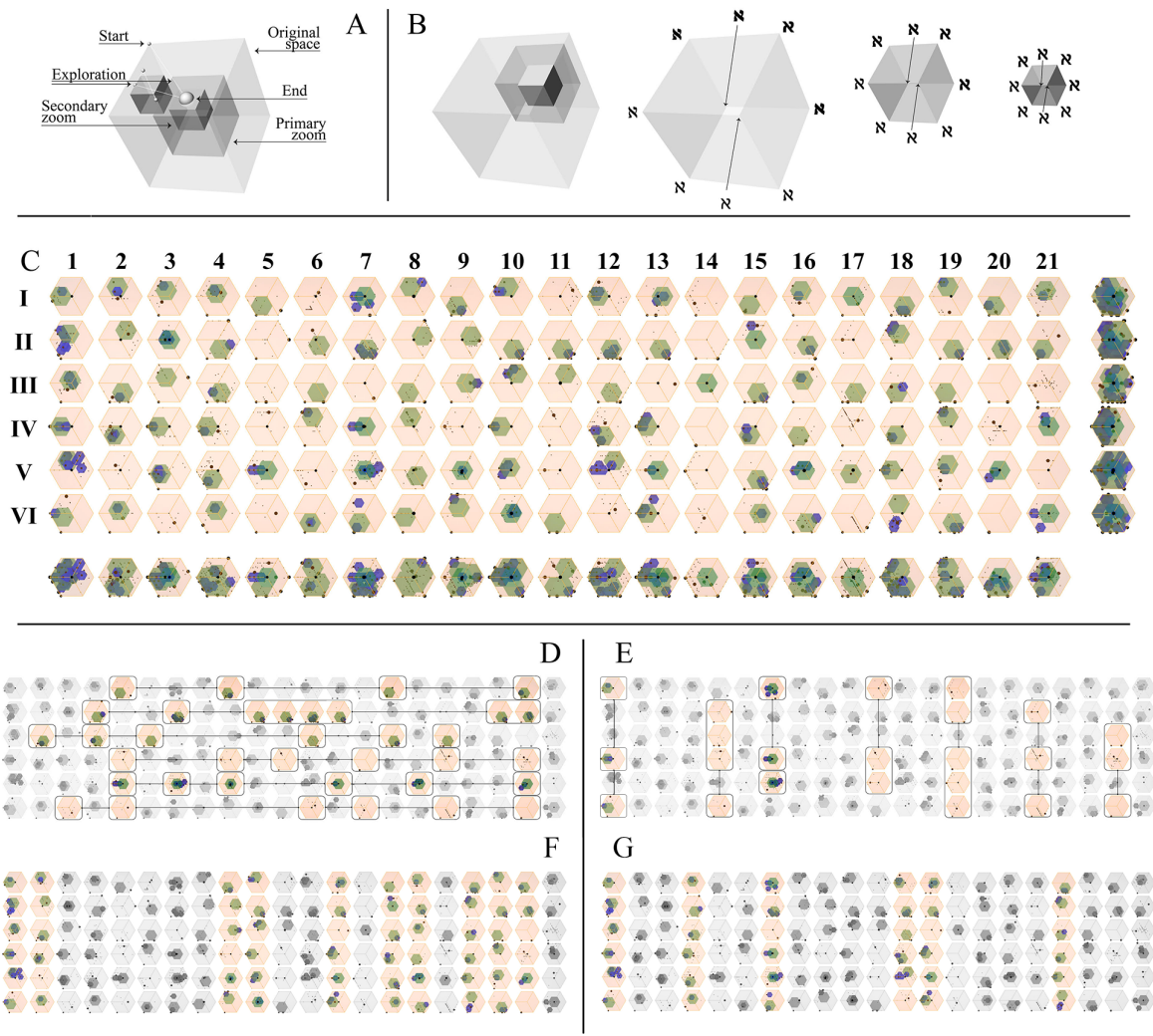

Figure 5: Study data visualization. (A) For each user selection, the Start instance, End instance – the downloaded font, The Exploration path and points of interest, and the Areas of Interest were recorded. (B) Zooming-in on areas of interest: each zoom-in action displays a subspace, adding instances to the display, growing in similarity. (C) Columns represent 21 users performing 6 tasks (rows). At the far right and far bottom, an accumulated display of selection patterns, for each task and user (respectively). (D) A high visual-correlation within tasks. (E) One-third of the participants presented a consistent recurring pattern of selection. (F) Half of the participants demonstrated convergence in their search. (G) One-quarter of the participants wandered excessively throughout the design-space. Note: Visualization is based on dimensional reduction.

***Selection Breadth.*** Half of the participants demonstrated convergence during their selection process for a given task. Their exploration was contained within the initial area of interest (Figure 5F, green medium-sized hexagons, containing small purple hexagons). Half of the final instance selections were found near their starting points (U1-3, U6-10, U13, U16, U20), and one-third of the users limited their search to an area close to their starting points (U1, U3, U6, U9-10, U18-20). These traversals can be characterized as depth-first search of the decision trees. One-quarter of the users wandered excessively throughout the design space, zooming-in on various areas in a given task (Figure 5G).

**Within-task Patterns.** A high visual correlation was present *across users* within tasks (Figure 5D). Three of the six tasks also presented a secondary cross-participant pattern that recurred between a smaller number of participants (tasks 3,4,6). Another correlation was observed between the starting point of search paths (a dot at an orange hexagon corner) and the given task; in tasks 3,4,5, and 6, the distinct majority of users started from the same instance.

**Visualization.** In the slider-based control interface, most users started each task by repeatedly moving back and forth between the two end points of each slider (as in 'blind selection' [28]): *"When there are sliders, I first go to the extreme to see what's available"* (U5). However, this did not result in the increased selection of extreme values. In contrast, this repetitive initial step did not occur while using *Inbetween*.

**Comparing the Interfaces.** When asked to compare the two interfaces, users noted the advantages of *Inbetween.* They preferred the simultaneous visualization of multiple options and extremes. Users felt that the full range of options was harder to reach in the control interface: *"Are you sure this is the same font?"* (U18), *"I like the sliders as they are more traditional, but I don't have the slightest clue how to get to all the options that the [Inbetween] showed me"* (U15). Users indicated that they kept their old habits using the sliders, overlooking the options they realized through *Inbetween*: *"With the sliders [in the control interface] I was aware of the values of every option, it's a habit [...]. I understood less the [Inbetween] interface but [using Inbetween] I reached much more interesting results"* (U8). The high-level partition lacking from the control interface was a frequent user comment: *"Where can I choose between serif and sans-serif here?"* (on the control interface, U10). However, at times, *Inbetween* was less intuitive: *"The sliders win for me, they give less results, but they work."* (U21), *"This zoom-in is just not intuitive"* (U2).

Overall, user feedback was very positive, and expressed interest in integrating the tool in their workflow: *"I feel as if [Inbetween] teaches me of all the nuances available, the simultaneous visuals are very useful"* (U16), *"I'd love to have this [...] for my workflow"* (U1), *"I'd like this as a [Adobe] plug-in"* (U21).

## 5 DISCUSSION AND FUTURE WORK

This research opens up a path towards defining and analyzing selection processes as a measure of personal style, by applying computer science tools. The data reveal initial indications of self-coherent *characteristics* in selection patterns, regardless of user expertise. We discuss the selection patterns and stylistic implications of our findings, consider the interface usability, and suggest future work.

*Inbetween*, a *minimal-bias* parametric design interactive tool, was created to investigate user selections. A user study was then employed to examine the *process* of user selection out of catalogs (rather than software usability) while using *Inbetween*. We hypothesized that we would find distinct differences between selection tasks and between users, especially with increasing years of experience. Our findings and analyses indeed showed consistency in selection patterns for each user, suggesting that such patterns are governed by the user's personal style. To enhance further selection analysis, future work should investigate these initial signs, acting along the suggested guidelines in creating minimal-bias tools.

This investigation required limitations of the (i) field, (ii) scope of users, (iii) space definition, and (iv) facilitation of interaction that might present visual bias. Even under these restrictions, we noted that experts did not present distinctly different selection patterns over novice users. While the lack of difference appears in the context of typeface selection, it is fundamentally opposed to our hypothesis.

**Scalability.** We hypothesized that we would find a selection pattern trend that correlated with experience. However, we found that while each user showed some self-coherent characteristics, there was no experience-related trend. To further inspect this somewhat counterintuitive phenomenon, future studies should span and scatter the experience of subjects. In addition, a high correlation in selection patterns *within* tasks was observed for each user. To this end, future studies and tool-making should examine task-related patterns.

**Shifting the Focus.** We recognize different types of selection patterns. Different manners of selection are easily noticeable, but not in the characteristics we thought they would be apparent. Some users were consistent in their selection patterns, repeating patterns and manners of exploration and selection throughout different tasks (e.g., U7), other users were somewhat inconsistent (e.g., U21), while yet others were notably inconsistent, utilizing a different pattern of search and selection in each task (e.g., U12). In other words, each user presented different characteristics in selection. While we initially tried to analyze and look at the same characteristic for all users, it became apparent that *selection characteristics* should be considered as a broad space incorporating a range of characteristics, rather than as a flattened aspect. In future works, we suggest defining a focused question regarding these findings, assuming each user is self-consistent within his/her unique characteristic combination. In an analogy from Linear Algebra, the user's *selection characteristics* can be defined as a linear combination of orthogonal vectors, each representing a different characteristic of selection patterns, possibly constructing a multidimensional space of selection characteristics.

**Visualization.** *Inbetween* eases the initial steps of the selection process thanks to visualization. First, the findings support the partitioning into initial categories. Second, visualization helped with initial orientation in the design space. Third, the repetitive initial step of moving between the two ends of each slider (in the control interface) was replaced by the initial visualization of the extreme values of *Inbetween*. *Inbetween* simultaneously presents the entire multi-dimensional design-space, in contrast to the slider control interface that enables manipulating only one parameter (i.e., dimension) at a time. Overall, user feedback regarding the simultaneous visualization of multiple options and extremes was very positive.

The following aspects need to be addressed to enable more comprehensive selection investigations.

**Interaction.** Given the multiple 'non-choosers' and 'minimal interaction' users (Section 4.3), the taken interaction approaches should be further investigated. It is possible that the sub-problem size also affected 'non-choosers' and 'minimal interaction' users. Notably, if user interactions were smoother and with less visual bias, the data produced could be cleaned of noise, allowing for closer investigation of selection patterns and personal style.

**Design Space.** Our design space definition allowed for three levels in the decision tree. Instances at the last level became too difficult to distinguish (Figure 5B). To enable a more detailed investigation of each selection process, future work should aim to span a deeper design space by presenting higher variance between instances.

**Platform.** For this research, we created an accessible and visually appealing web tool. However, user feedback suggested the need for a tool that could be more integrated within the designer workspace.

***Utilizing the Chaos.*** A recurring element arising in this analysis was the need for a highly adaptive tool. User interfaces generally try to control chaotic and intuitive approaches to facilitate processes and interactions. An alternative approach [81] allows the chaos, leveraging it within an enabling environment. As we have shown, each user employs a unique approach and process, not shared even within a level of mastery. Thus, an ideal tool should react to the user-specific pattern, i.e., the self-consistent unique combination of characteristics.

## 6 CONCLUSIONS

In this paper, we presented *Inbetween*, an interactive *minimal-bias* parametric-design catalog for visual font selection. The catalog was used to examine visual selection patterns, as a means to study *style*. *Inbetween* visualized the multi-dimensional design space, enabling interactive visual exploration and selection. Our user study examined the *process* of user selection while using *Inbetween*. The 21 participants were practicing graphic designers, without prior exposure to parametric design. We followed and recorded their visual selection processes, which were then formalized and analyzed as traversals of decision trees. In addition, usability feedback about *Inbetween* was collected.

The contributions of this work include (i) catalog creation guidelines aspiring minimal visual bias and (ii) *Inbetween*, a visual selection catalog following the guidelines, enabling the investigation of style.

Our findings revealed self-coherent *characteristics* in selection patterns. We also found a high correlation in selection patterns *within* tasks. These user-specific characteristics and selection patterns reflect the user's personal style. This work demonstrated the application of computer science tools to the study of style. While this study focused on typographic interaction, the findings can be generally applied toward definition of selection processes as a measure of personal style.

## ACKNOWLEDGMENTS

We would like to thank Daniel Grumer, Kobi Levi, and all the design studios that volunteered their time for our user study.

## A  APPENDIX

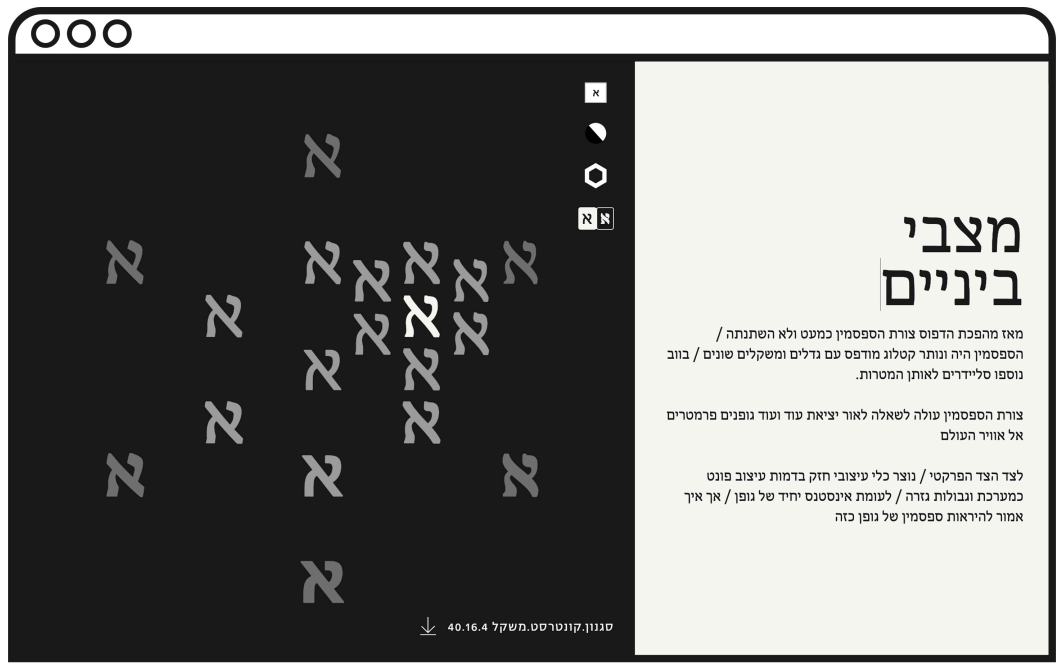

A screenshot of the original Hebrew version of our *Inbetween* interface used in the user study.